\documentclass{article}
\usepackage{amsmath}
\usepackage{hyperref}
\usepackage{epsfig}
\usepackage{graphicx,caption}
\usepackage{comment}
\usepackage{bm}
\def\dsize{\displaystyle}
\def\E{\mathop{\rm E}\nolimits}
\def\Ro{\mathop{\rm Ro}\nolimits}

\def\Ra{\mathop{\rm Ra}\nolimits}

\def\Pr{\mathop{\rm Pr}\nolimits}

\begin{document}
\begin{center}{
\bf EVOLUTION OF THE INNER CORE OF THE EARTH: \\  CONSEQUENCES FOR GEODYNAMO }

\end{center}

\begin{center}{M. Yu. Reshetnyak  \\ 
\vskip 0.5cm

 Institute of Physics of the Earth of RAS,
  Moscow, Russia, 
  
     Institute of Terrestrial Magnetism, Ionosphere and Radio Wave Propagation of RAS, Moscow, Russia, 
 m.reshetnyak@gmail.com}
\end{center}

\begin{abstract}
           Using    models of the Earth's core evolution  and the length of the day observations the change of the dimensionless geodynamo parameters is considered. The evolutionary model  includes  cooling of the liquid adiabatic core,  growing solid core, and the region in the outer part of the core with a  subadiabatic temperature gradient. The model covers time period 4.5Ga in  the past  till 1.5Gy to the future, and   produce  evolution of the   energy sources of the thermal and compositional convection, spatial scales of the convective zone. These quantities are used for Ekman, Rayleigh and Rossby numbers estimates. So far these numbers determine regime of the geomagnetic field generation,  we discuss evolution of the geomagnetic field over Earth's evolution.     \end{abstract}


\section*{Introduction}
   
                       Dynamo theory is the most promising  candidate to explain existence of the geomagnetic field observed at the surface of the Earth \cite{Roberts:King}.                      
                           If the direct three-dimensional modeling of the magnetic field generation over the times comparable to the age of the Earth ($4.5$Gy) is still beyond the modern computer capacities,  because of the wide range of the temporal-space  scales of the MHD turbulence in the core, then   study of influence of the evolution of the core on the geodynamo parameters 
           is already quite realistic  task  \cite{DRISCOLL200924}.
         
            The straightforward  application of the   evolutionary scenarios to  the geodynamo models is complicated by the different representation  of parameters. Thus, the evolutionary models produce evolution of the inner core and subadiabatic region boundaries, heat fluxes. In other words the models give the behavior of the physical quantities. 
   On the other hand, the geodynamo models have deal with a set of the dimensionless parameters, which are very useful for analysis of the physical state in the core. 
      As a result translation from  one language to another is needed.

     Below we repeat the well-known and accepted by the geodynamo community modeling  of  the thermal evolution model of the core, originated to the pioneer papers 
             \cite{GubbinsMasters1979, GubbinsThomson1982,   LABROSSE19971, STEVENSON20031, LABROSSE2003127} and references therein, which include  cooling of the liquid core, evolution of the geometry of the liquid core, concerned with the growth of the inner core and subadiabatic region, change of the energy sources. 
                 So far, the Earth's   angular  velocity, which contributes to the dimensionless numbers,  evolves on the geological times due to the tidal forces, information on  variations of the length of the day from observations was used as well \cite{varga1998}. As a result we get  evolution of the Ekman, Rayleigh, Rossby numbers, the critical Rayleigh number for the thermal and compositional convection starting from the origin of the liquid core about 4.5Ga,  and extrapolate it to a close, on a geological scales, future 1.5Gy. We also  discuss the known paradox \cite{Olson2013}, concerned with the fact that compositional convection, started with the origin of the inner core,  should change  the geomagnetic field generation in the core.
                   
   \section{Thermodynamic model of the  core}
            Following \cite{GubbinsThomson1982,LABROSSE19971, LABROSSE2003127} we consider scenario of the Earth's evolution, where soon after the end of the  accretion process, the Earth's core of radius $r_b$   was fully convective. Then,   it cooled due to the thermal flux density $q_b$ at the core-mantle boundary (CMB) $r=r_b$, and as a result, depending  on the amplitude of $q_b$,  two regions  could  appear: the solid inner core ($0\le r\le c$, region I) and   subadiabatic layer in the outer part of the core ($r_1\le r\le r_b$, region III). The rest convective part of the core $c\le r\le r_1$ here and after is denoted as region II.

Radial distributions of density  $\rho(r)$, pressure $P(r)$ and gravity  $g(r)$   satisfy to the hydrostatic balance equations:
\begin{equation}\dsize
          \nabla P=-\rho g, \qquad \dsize
         g(r)={4\pi G\over r^2} \int\limits_0^r \rho(u) u^2\, du,
         \label{c1}
\end{equation}
with  $G$ the gravitational constant.
    To close system of equations for  $(P,\,  \rho,\, g)$  the logarithmic equation of state  is used:
\begin{equation}
          P= K_\circ{\rho\over\rho_\circ}  \ln {\rho\over\rho_\circ},                
        \label{c3}
\end{equation}
  where  $K_\circ$,  $\rho_\circ$ are  incompressibility and density at zero pressure, respectively.
      The optional in the model  jump of the density, observed at the surface of the inner core, and which effect on the evolution of the core is quite  small, is introduced as follows:
\begin{equation}\dsize
 \rho(r) = 
  \begin{cases} 
    \rho(r),   & \text{if } r \le c \\
         \rho(r)-\delta\rho,   & \text{if } r> c.
   \end{cases}
        \label{c5}
\end{equation}
  Eqs(\ref{c1}--\ref{c5}) with given $c$ can be  solved numerically. Then, with known $(P,\,  \rho,\, g)$, adiabatic temperature profile can be derived:
 \begin{equation}\dsize
           T_{ad}(r)=T_c(c)\, exp\left(\dsize -\int\limits_c^r {\alpha(u) g(u)\over C_p} \, du \right),
         \label{c6}
\end{equation}
   where  $T_c(c)$ is the temperature at  $r=c$,   thermal expansion coefficient
\begin{equation}\begin{array}{l}\dsize
             \alpha(r)={\gamma C_p\rho_\circ   \over K_\circ\Big( 1+ \dsize  \ln{\rho\over\rho_\circ} \Big) },
                 \end{array}
        \label{c7}
\end{equation}
     with  $C_p$ specific heat, and $\gamma$ for Gr\"uneisen parameter.

 If the inner core is still absent, $c=0$, then  $T_c(c)=T_\circ$, where  the temperature in the center of the Earth $T_\circ$  
   can be found from the heat balance equation:
    \begin{equation}\dsize
              4\pi  r_1^2 q_1= -  4\pi  \int\limits_0^{r_1} \rho C_p {\partial T_{ad}  \over \partial t}   r^2\, dr=    
              -{\partial T_\circ S \over \partial t}, \qquad  S=4\pi\int\limits_0^{r_1} \rho C_p exp\left(\dsize -\int\limits_0^r {\alpha g\over C_p}\right)     r^2\, dr,
         \label{c8}
\end{equation}
   with   $q_1$ for    heat flux  density  at $r_1$. 
 The growth of the inner core starts, when temperature of the liquid core  is equal to the  temperature of solidification:
     \begin{equation}\begin{array}{l}\dsize
                 T_s(r)=T_s^\circ \left(\dsize {\rho(r) \over \dsize   \rho(c)}\right)^{2(\gamma-{1\over 3 })}   ,
                      \end{array}
        \label{c10}
\end{equation}
          where $T_s^\circ$  is the temperature of solidification  in the center of the Earth.
                        Solidification process starts in the core's center, i.e.  $T_c=T_\circ=T_s^\circ$,  $r=c=0$.
              Then,  for  $c>0$, $ T_s$   defines adiabatic temperature at the boundary $c$ in (\ref{c6}):  $T_c(c)=T_s(c)$.
 
               Position of the inner core boundary  $c$  can be derived from the heat flux equation:
  \begin{equation}\begin{array}{l}\dsize
             r_b^2 q_b-c^2 q_c=   \dot c \Big(        c^2   (P_L+P_G)  +  P_C \Big),
          \end{array}
        \label{c11}
\end{equation}
  where on the left side is the  total heat flux in the region II, and on the right one  are the cooling sources, and the dot over $c$ stands for the time derivative. 

   The latent heat source is defined as 
  \begin{equation}\begin{array}{l}\dsize
    P_L(c) = \rho(c)\delta S\, T_s(c),
           \end{array}
        \label{c12}
\end{equation}
with $\delta S$   entropy of crystallization.

 Estimate of the release of the  gravitational energy due to the growth of the inner core has the form   \cite{Loper1984}:
 \begin{equation}\begin{array}{l}\dsize
            E_G=   {2\pi \over 5}  G M_\circ \delta \rho {c^3\over c_b} \left(1-\left({c\over r_b}\right)^2 \right),
          \end{array}
         \label{c13}
\end{equation}
     with   mass of the core   $\dsize M_\circ={4\over 3}\pi\int\limits_0^{r_b}\rho(r) r^2\, dr$  constant in the model.
Then  it leads to
  \begin{equation}\begin{array}{l}\dsize
     \dot      E_G= 
        P_G\dot c, \qquad P_G={12\pi\over 5} {G M_\circ\delta\rho\over r_b} c \left(1-{2c^2\over r_b^2}\right).
          \end{array}
         \label{c14}
\end{equation}
 The main term, concerned with adiabatic cooling, has the form:
 \begin{equation}\begin{array}{l}\dsize
            P_C =
                   -    \int\limits_c^{r_1}\dsize \rho C_p {\partial T_{ad}  \over \partial c}      r^2\, dr,\qquad dc\equiv dr,
           \end{array}
        \label{c15}
\end{equation}
   with $q_c$  heat flux   density through the boundary $c$.  
      Eq(\ref{c11}) was resolved with respect to   $\dot c$ and then integrated in time. This defines evolution of the inner core boundary $c$ in time.

From condition of continuity of the temperature at the boundary $c$, follows that  $T_s(c)$  is the boundary condition 
  for the thermal-diffusion equation in the region  $0<r<c$,   I,  with a moving boundary  $c(t)$ \cite{KUTLUAY1997135}:
    \begin{equation}\begin{array}{l}\dsize
            {\partial   T\over \partial t}=k \Delta T,
                                 \end{array}
         \label{c16}
\end{equation}
where  $k$ is the thermal diffusivity. 
  The second boundary condition in the center  $r=0$ is  
      $T'=0$, where $'$ is a derivative on $r$. The joined system  $(\ref{c1}-\ref{c16})$ defines evolution of the fields in the regions I and II.

             If  the      adiabatic heat flux density  $\dsize q_{ad}(r)=-\kappa T'_{ad}(r)$, with thermal conductivity  $\kappa=k\rho C_P$,  becomes  larger than the  heat flux density  at  the outer boundary $r_b$: $\dsize q_{ad}(r)<\left( {r_b\over r }   \right)^2 q_b$, the subadiabatic stably stratified thermal   region III  develops at the outer part o the core, where the  heat flux density is smaller.  
       The temperature profile in  the region III can be derived from  Eq(\ref{c12}) with the moving boundary  $r_1(t)$, and two boundary conditions:
      $T(r_1)=T_{ad}(r_1)$ at the inner boundary,  and given heat flux density $q_b(t)$   at the outer boundary $r_b$. 
         In the general case Eqs(\ref{c1}--\ref{c16}) in regions  I--III  are solved numerically, using iterative methods with under-relaxation method to provide  numerical stability.

      \section{Evolution of the boundaries $c$ and $r_1$}
      We consider two regimes  similar to that ones in  \cite{LABROSSE19971}, proposed  by the authors, to get the 
     realistic size of the inner core at the present time. These regimes differ in decrease of the heat flux density 
    $q_\beta$ in the linear dependency $q_b(t)=q_b(0)-q_\beta t$ for the heat flux intensity at CMB.  
     Let at $t=0$   $q_b(0)=0.075$ W/m$^2$,  then in the case $A$  $q_\beta=10^{-19}$ W/(s\,m$^2$), and  $q_\beta=3\, 10^{-19}$ W/(s\,m$^2$)   in the case $B$.
      These values correspond to decrease of the heat flux density of 4.2\% and  12.6\% during the first  1Gy, correspondingly.
             Eqs(\ref{c1}--\ref{c16}), with parameters summarized  in  Table~1,  were solved numerically for the time interval 6Gy. This time interval corresponds   to 4.5Ga to the past,  starting from the origin of the liquid core, till  1.5Gy to  the future.
               \begin{table}
    \caption{ }
   \begin{center}
   \begin{tabular}{c|c|c}
   \hline     
      Parameter  &  Denotation  & Value \\
 \hline     
       Gravitational constant & $G$   & $\rm 6.6873\, 10^{-11}\, m^3/(kg\,s^2)$  \\
     Thermal diffusivity      & $k$  & $\rm 7\, 10^{-6}\, m^2/s$  \\
       Kinematic viscosity  & $\nu$  & $\rm 10^{-6}\, m^2/s$  \\
 Light element diffusivity    & $\lambda$  & $\rm 10^{-9}\, m^2/s$  \\          
   Coefficient of chemical expansion   & $\beta$  & $1$  \\
  Gru\"uneisen parameter     & $\gamma$  & $1.5$  \\
    Core radius       & $r_b$ & $\rm 3480\, km$  \\
 Entropy of crystallization      & $\delta S$ & $\rm 118\, J/(kg\, K)$  \\
   Density at zero pressure    & $ \rho_\circ$  & $\rm 7500\, kg/m^3$  \\
         Density jump at ICB & $\delta \rho$  & $\rm 500\, kg/m^3$  \\
  Temperature of solidi¢cation at the cente     & $T_s^\circ$ & $5270\,\rm  K$  \\
       Initial temperature in the center of the core      & $ T_\circ$ & $6000\,\rm  K$  \\
   Incompressibility of the core & $K_\circ$ & $ 4.76\, 10^{11}\, \rm   Pa$ \\
        Specific heat     & $C_p$ & $\rm 860\, J/(kg\,K)$  \\
   \hline     
 \end{tabular}
\end{center}
\end{table}

\begin{figure}[h!]
                     \vskip -2.6cm
  \def \ss {7.5cm}
    \hskip 5cm \includegraphics[width=\ss]{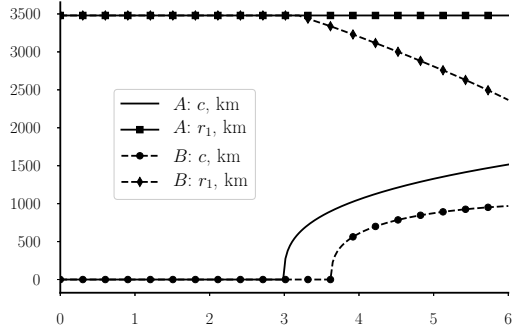}
           \vskip -2.7cm
         \caption{Evolution of the inner core boundary $c$ and boundary of the stably stratified layer $r_1$ in  cases   $A$ and $B$.
 } \label{fig_1}
\end{figure}

    Evidently that cases $A$ and $B$ differ in  evolution of the boundaries  $c$ and $r_1$, see Fig.\ref{fig_1}.
   In the case  $A$ region III is absent  during the whole time interval, and the inner core  appears  earlier ($t=3$Gy) than for $B$ ($t=3.6$Gy). 
     In the case $B$ firstly appears region III ($t=3.2$ Gy), and after that  the inner core (I). The velocity  $\dot r_1$ in case $B$ is approximately constant in time and greater than  $\dot c$. The age of the inner core in cases $A$ and $B$ is $1.5$Ga and $0.9$Ga, correspondingly.
   In the case $A$  and $B$ at the present time   the  inner core radius $c$   is equal to  $1230$km and $810$km, correspondingly. Note, that seismological estimate of $c$ is 
      $1220$km. In the case $B$, where the heat flux at CMB decreases faster, the inner core radius is smaller.
     
                        \section{Evolution of the length of the day} So far  convection and magnetic field generation in the liquid core are closely related to the angular rotation velocity $\Omega$  of the Earth, evolution of $\Omega$ in time should be also considered. Due to the tidal forces Earth's length of the day (LOD)  $\dsize T_{\rm d}=   {2\pi \over \Omega}$   increases in time. The estimate of this increase  over the last three centuries  is   $(2.0\pm0.2)$ms cy$^{-1}$. This process  can be described analytically, using astronomical predictions on the evolution of the Earth's orbit, and the results are in agreement with observations.  However astronomical methods can not be extrapolated to the geological times, because they do not include evolution of the Earth itself. This is the reason why  the new palaeontological and palaeosedimental observations  should be used, see review \cite{varga1998}.
                         The relevant information comes from fossils (bivalves, brachiopods and corals) of the Phanerozoic (time interval from  542Ma to the present time), from stromatolites mainly of the Proterozoic (from 2500Ma to the Phanerozoic), and from palaeodeposits of the Proterozoic. The main idea of these approaches is that one can resolve as the daily marks of the growth of these organisms, as well as the annual, and even seasonal ones.
                                                     Then  finding the number of days per year is the trivial  procedure. To the moment information on LOD is available for the last 2.5Gy.  
                         To my knowledge such a long time series of LOD were never used in the geodynamo 
                          theory.  Here  we  extrapolate data from   \cite{varga1998} to the last 6Gy in the following form:
   \begin{equation} T_{\rm d}(t)=
     \begin{cases}
           21.435+0.974(t-4.5) , & 0\le t\le 3.86, \\ 
       24+4.98\, (t-4.5) , & 3.86\le t\le 6,  
       \end{cases}
       \label{sys1}
   \end{equation}
      where LOD $T_{\rm d}$ is measured in hours, and $t$ in Gy. The present time corresponds to $t=4.5$. There is the break at $t=3.86$, when  $T_{\rm d}$ starts to decrease faster in factor 5. However the certain estimates of accuracy of the data hardly can be done, authors of \cite{varga1998} demonstrate that the break is statistically significant. One of the possible explanation of such a break, considered by authors,  is the change of the plate tectonics of the Earth.

          \section{Dimensionless numbers} The  Ekman number $\dsize \E={\nu\over 2 \Omega D^2}$, with  $D$ typical scale of convection, characterizes ratio of the viscous and Coriolis forces. 
      Definition of the scale $D$ depends on the model of convection. For the thermal convection, denoted by index $\rm T$, the stably stratified layer III is excluded, and    $D=D_{\rm T}=r_1-c$. For the compositional convection (index $\rm C$), the full convective liquid core is considered $D=D_{\rm C}=r_b-c$. If region III is absent, then $D_{\rm T}=D_{\rm C}$.
       
        Evolution of $\E_{\rm T}$  is presented in Fig.2a. There is the slight increase of $\E_{\rm T}$ before the inner core origin, caused by  change  of $T_{\rm d}$ (\ref{sys1}).
    After  regions  I and III have developed,  evolution  of $\E_{\rm T}$ is controlled by the scale of convection $D_{\rm T}(t)$.  
     For compositional convection for the case $A$ $\E_{\rm T}=\E_{\rm C}$ because $D_{\rm T}=D_{\rm C}$. In the  regime $B$ the growth of  $\rm E_{\rm C}$ is less, because   $D_{\rm C}$ does not include  $r_1(t)$. As a result, curves do not intersect. 
  
\begin{figure}[h] 
   \def \ss {6cm}
\begin{minipage}[t]{0.5\linewidth}
            \vskip -2.0cm
   \hskip -0.3cm \includegraphics[width=\ss]{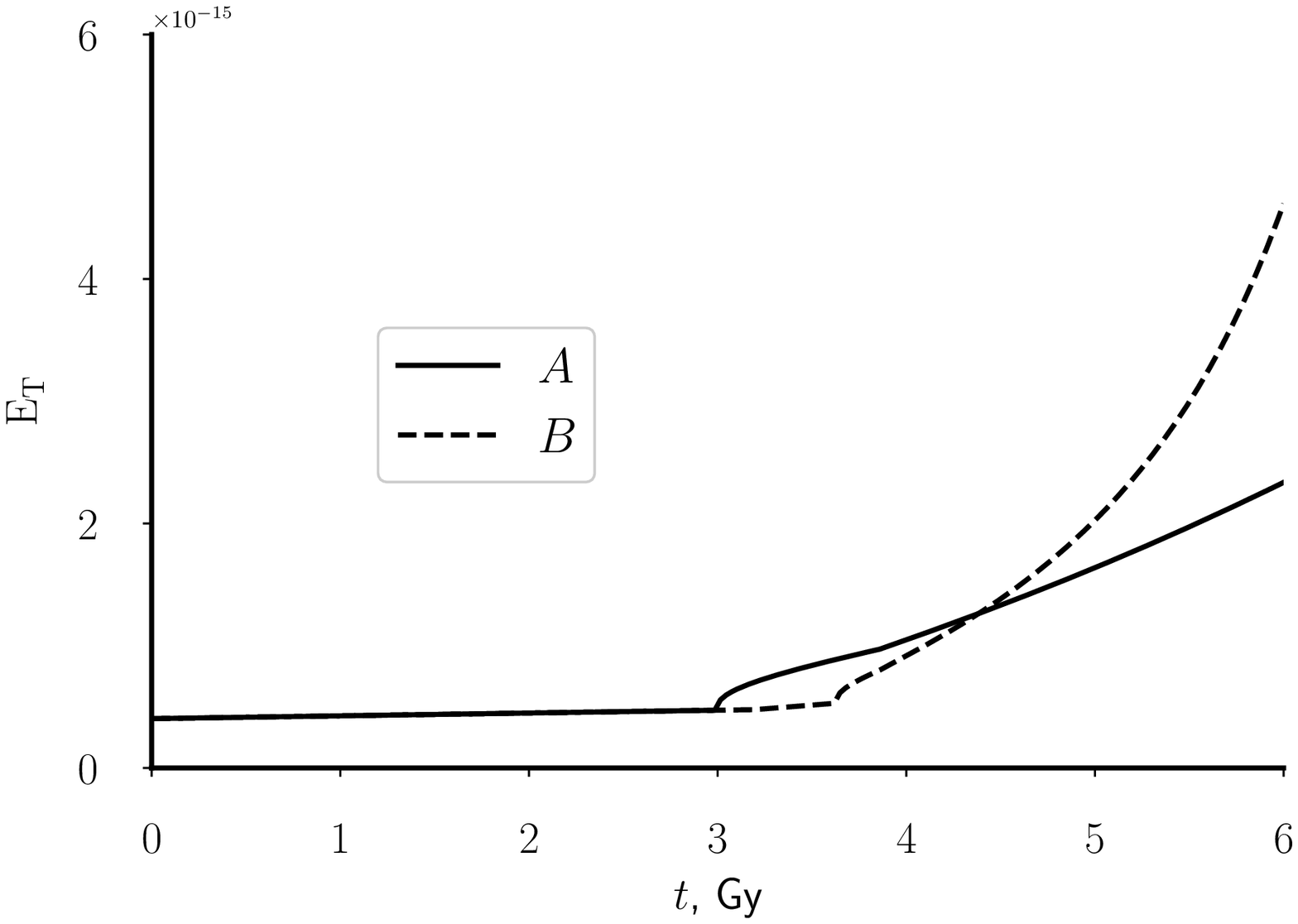}
\end{minipage}
\begin{minipage}[t]{.5\linewidth}
                     \vskip -2.1cm
    \hskip -0.1cm \includegraphics[width=\ss]{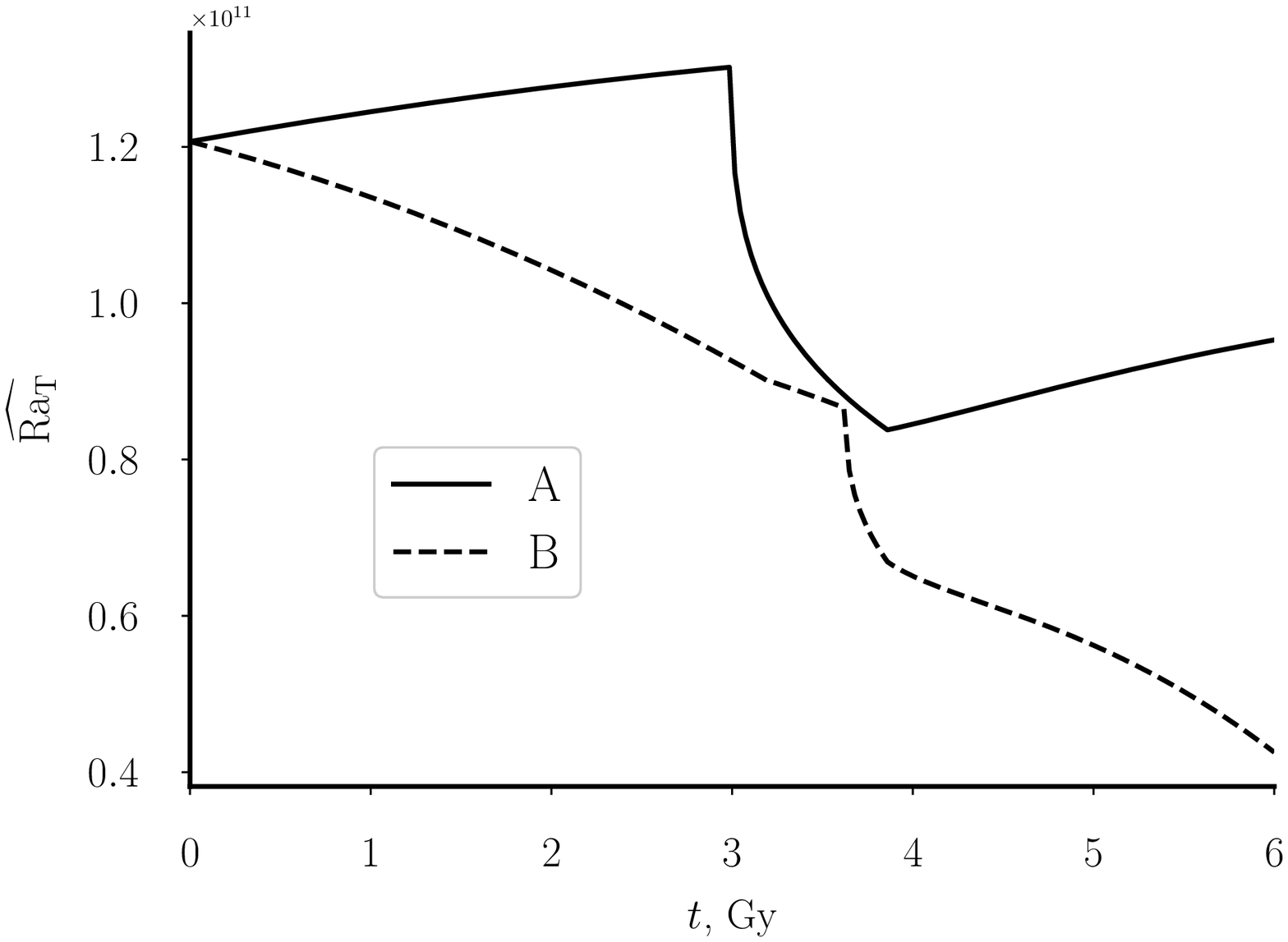}
\end{minipage}
       \vskip 0.6cm
 \begin{minipage}[t]{.5\linewidth}
             \vskip -4.6cm
  \hskip 5cm \includegraphics[width=\ss]{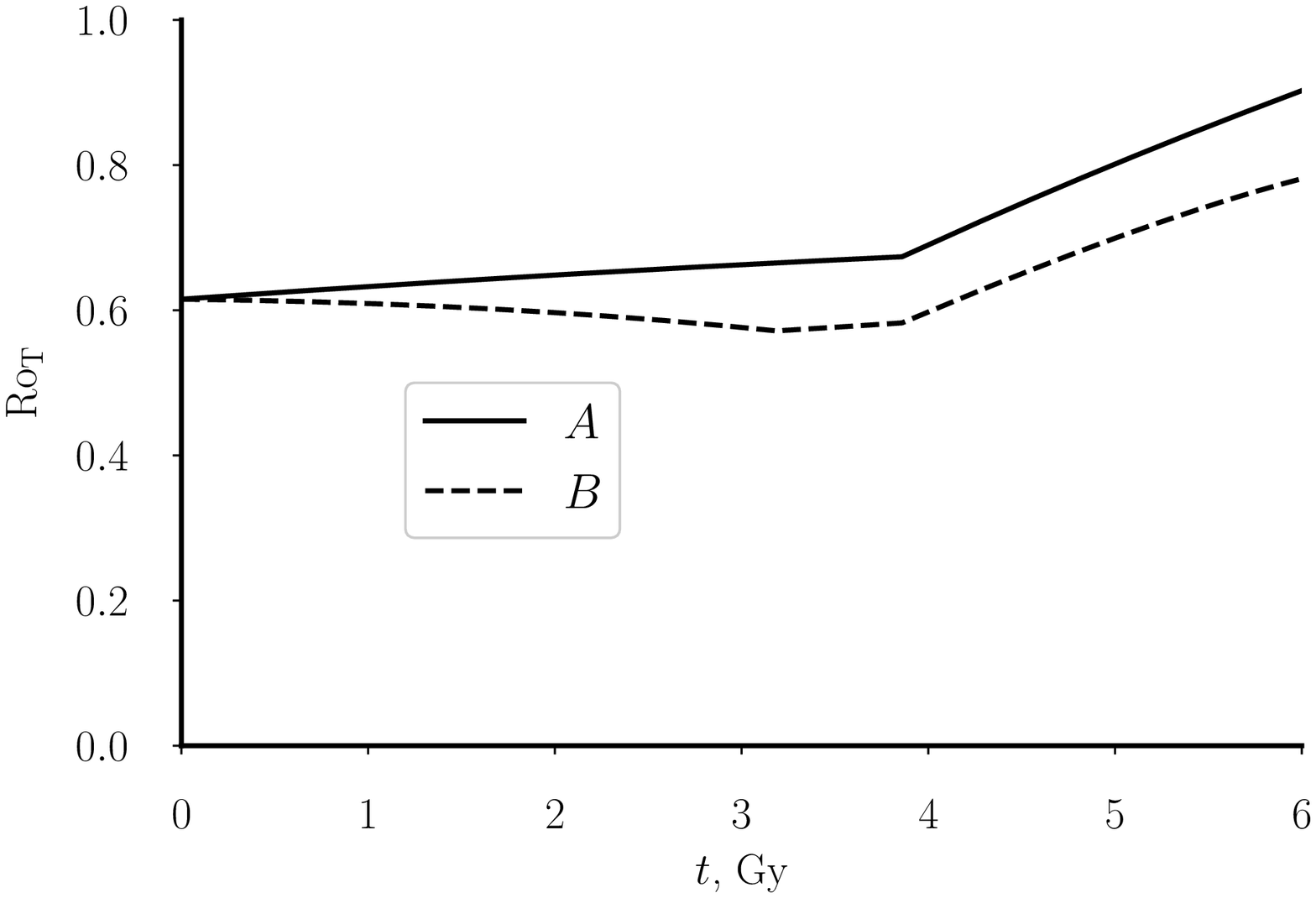}
 \end{minipage}
                \vskip -9.5 cm 
                  \hskip 5cm (a)  \hskip 8cm (b) 
           \vskip 4.0cm 
                  \hskip 11.0cm (c) 
                          \vskip 2.6cm
               \caption{ Evolution of the Ekman number $\E_{\rm T}$ (a), normalized Rayleigh number $\rm \widehat{Ra_T}$ (b), and the Rossby number $\rm Ro_{\rm T}$ (c)
     for  $A$ and $B$.} \label{fig_2}
\end{figure}

     The measure of  the heat sources intensity is the Rayleigh number: 
       $\dsize \Ra_{\rm T}=\dsize {\alpha g q_{r_1} D_{\rm T}^4\over \kappa k   \nu  }$, with $\dsize q_{r_1}(t)=   \Big({r_b\over r_1}\Big)^2 q_b(t)$. It is instructive to consider  $\dsize \Ra_{\rm T}$  in units of its critical value  $\dsize \rm Ra^{cr}$:   $\rm\widehat{\Ra}=\Ra/\Ra^{cr}$. For  $\E_{\rm T}\ll 1$    $\dsize \rm Ra^{cr} \sim {\Pr^{4/3}\over (\Pr+1)^{1/3} }   E_T^{-4/3}$, and  in the case 
           of the thermal convection, where the Prandtl number  $\dsize \Pr={\nu\over\kappa}\approx 0.1$  is small,  $\dsize \rm Ra^{cr}\sim {\Pr^{4/3} }   E_T^{-4/3}$.  Before the inner core appears,  $\rm \widehat{Ra_T}$, due to evolution of 
          $q_b(t)$ and $\Omega(t)$,   evolves   slightly different in cases $A$ and $B$, Fig.~\ref{fig_2}b.  However, after the start of the inner core solidification, when $\dot c$ is large, there is sharp decrease of   $\rm \widehat{Ra_T}$ in the both cases. Duration of such decrease is limited by the time moment, when the length of the day changes in the model ($t=3.86$Gy). After that $\rm \widehat{Ra_T}$ continues to grow in the case $A$, and  decrease in $B$. Note, that if the origin of the inner core would coincide with the break in dependence (\ref{sys1}), then the jump in   $\rm \widehat{Ra_T}$ will disappear. This fact  can be an implicit justification that the break in (14)  corresponds to the inner core emergency.
   The Rossby number $\Ro$ is the ratio of $T_{\rm d}$ to the typical convective time. The larger is $\Ro$ the more negligible is the Coriolis force in the system. The value of $\Ro$ can be derived from the following scaling-law  $\dsize \Ro_{\rm T}= \left({\Ra_{\rm T}\,  \E_{\rm T}^2}\right)^{0.41}$ \cite{ChAu2006}. 
        The behavior of  $\Ro_{\rm T}$ is similar in  the both cases $A$ and $B$, see Fig.~\ref{fig_2}c: the  slow change till the break in $T_{\rm d}$, and after that increase for 50\% during 2Gy. For the present time one has to divide 50\% on factor 4. Summarizing, we conclude that increase of $\Ro_{\rm T}$, based on the thermal convection model, after the origin of the inner core is quite small.

     For the  compositional convection the following estimate of the Rayleigh number is used  \cite{olson2008course}:
   \begin{equation}
             \dsize \Ra_{\rm C}=\dsize {\beta g_0\dot\chi_\circ  {D}_{\rm C}^5\over
    \lambda\nu^2},
         \label{sys10}
\end{equation}
     with $\lambda$ for the  light element diffusivity,  $\dsize\beta=-{1\over \rho}{\partial \rho\over\partial \chi}$ coefficient of the chemical expansion, and 
 time derivative of the 
  light element concentration   $\chi_\circ  $  is related to  $\dot c$ as 
          \cite{olson2008course, DRISCOLL200924}:
   \begin{equation}
                \dot\chi_\circ= { 3\chi_\circ\over 1-\chi_\circ  }
               {c^2\dot c \over r_b^3  - c^3}.
        \label{sys11}
\end{equation}
   In the limit of  $\chi_\circ\ll 1$, and $c\ll r_b$, Eq(\ref{sys11}) has exact solution: 
    \begin{equation}
                          \chi_\circ(t) =  C \, exp\left( 3{ \left( c(t)\over r_b\right)^3}\right),  
          \label{sys12}
\end{equation}
        where     $C$ is the constant, defined by the initial condition that in the present time 
       $\chi_\circ\sim 0.1$  \cite{DRISCOLL200924}. 
   Using calculated dependence  $c(t)$ one estimates  that  $\chi_\circ$ in cases $A$ and $B$ is in the range of $[0.89\div 0.11]$,   $[0.97\div 0.1]$, correspondingly. 
      
    Estimate of $\rm \Ra_{\rm C}$  gives value three orders of magnitude  larger than for the thermal convection \cite{olson2008course}, however this difference for the normalized values  $\widehat{\Ra}$ is only factor 20, see Fig.~2b and 3b. It is because for the compositional convection $\dsize \Pr={\nu\over \lambda}\sim 10^3$, and $\Ra^{\rm cr}\sim \Pr\E^{-4/3}_{\rm C}$.
       The time dependence of $\rm \widehat{\Ra_{\rm C}}$ is similar to that ones for  the thermal convection regime in  Fig.~2b.
        Again, if the origin of the inner core coincides with the break in  $T_{\rm d}$, the jump in curve  $\widehat{\Ra_{\rm C}}$ disappears.

     By analogy with the thermal convection regime, estimate of the Rossby number for the compositional convection $\dsize\rm \Ro_C= \left({\Ra_{\rm C}\,  \E_C^2}\right)^{0.41}$ 
   is plotted in Fig.~3b.  Firstly note, that this estimate leads to the quite large $ \dsize\rm \Ro_C\gg 1$, which hardly corresponds to the geostrophic state in the liquid core. Possibly, the additional  normalization for $\rm \Ra_C$ is needed. Let focus our attention at the time behavior of the curves, which   differs also in the cases $A$ and $B$: for $A$ contribution of $\dot\E>0$ is more  essential  than of $\dot\Ra_C<0$, and for $B$ situation is quite opposite. It follows  from 3D simulations \cite{ChAu2006}  that  Rossby number controls the  frequency of  the geomagnetic  reversals: the smaller is $\rm Ro$ the faster geomagnetic dipole changes its polarity. From this point of view the case $A$ looks more attractive, because of the general increase of the number of reversals after the Cretaceous superchrone.
          \begin{figure}[h!] 
   \def \ss {6cm}
                       \vskip .5cm
 \begin{minipage}[t]{0.5\linewidth}
    \vskip -2cm
     \hskip -0.5cm \includegraphics[width=\ss]{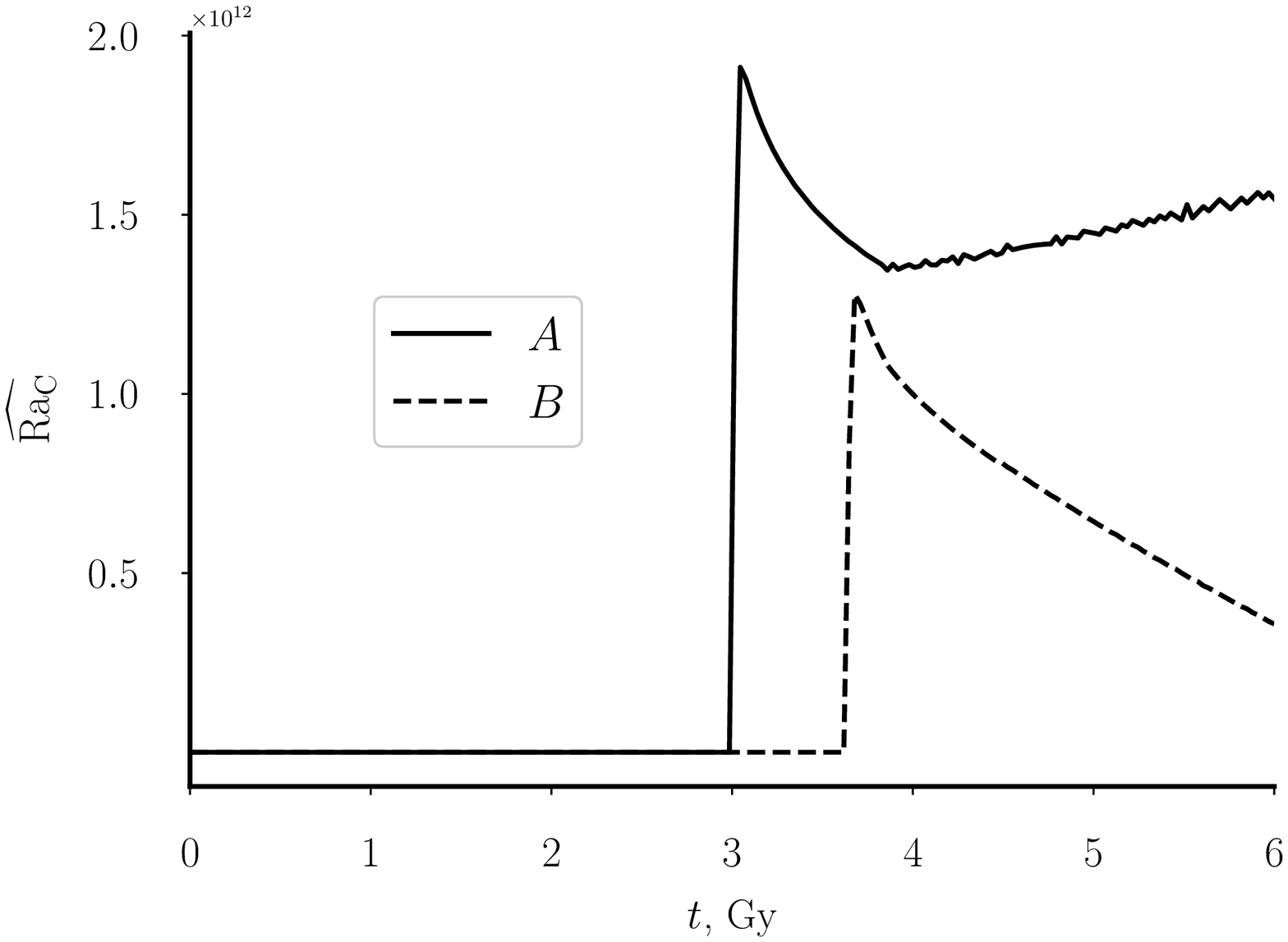}
\end{minipage}
 \begin{minipage}[t]{.5\linewidth}
               \vskip -1.9cm
      \hskip -1cm \includegraphics[width=\ss]{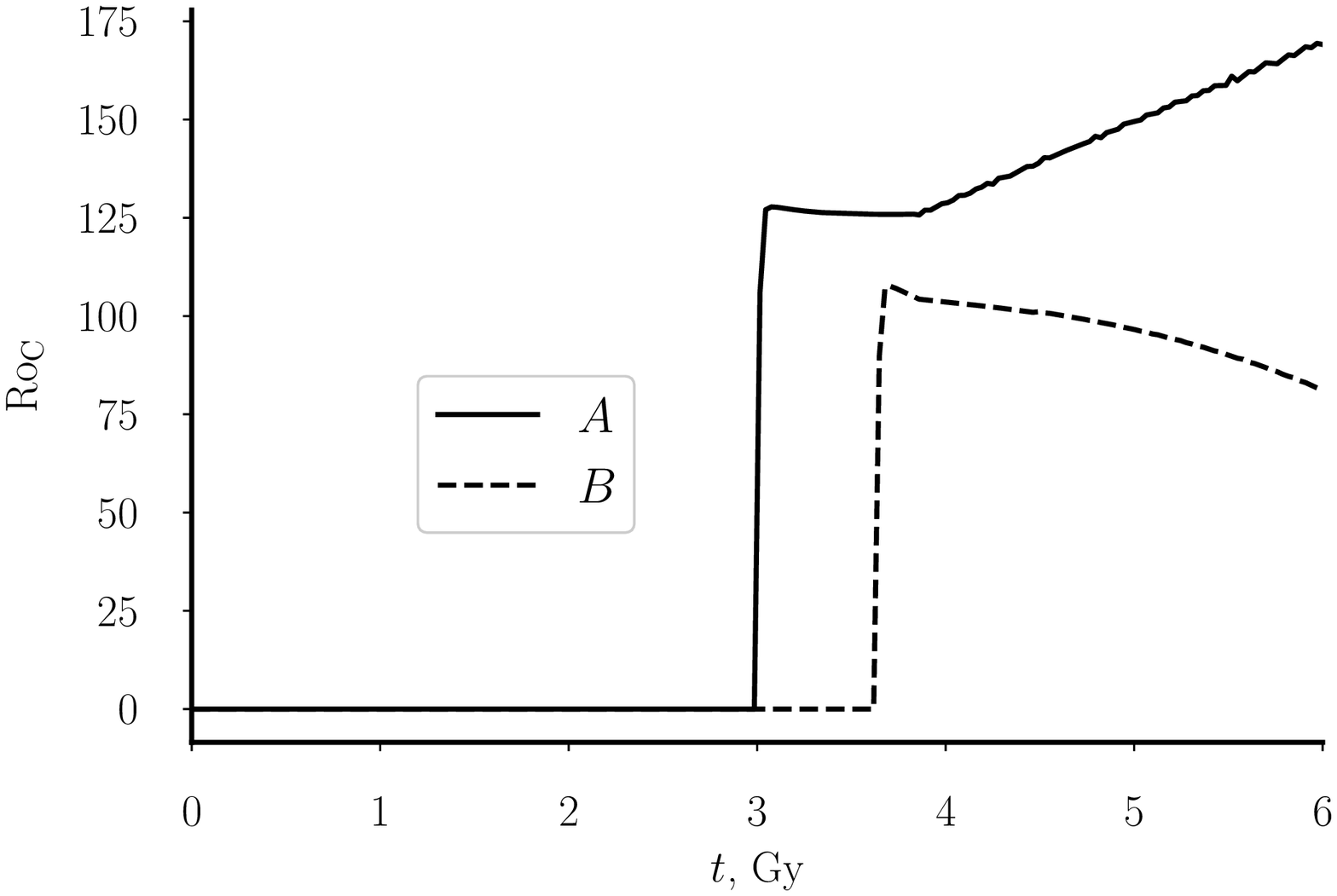}
\end{minipage}
                 \vskip -6.0 cm 
            \hskip 5cm (a)   \hskip 7.8cm (b) 
                     \vskip 3.5cm
  \caption{ Evolution of the normalized Rayleigh number $\rm \widehat{Ra_C}$ (a), and the Rossby number $\rm Ro_C$ (b)
   for the cases  $A$ and $B$.
} \label{fig_3}
\end{figure}
  \section{Conclusions}
          The evolutionary models predict existence of the young inner core $\sim 1$Ga. 
 Switch on of the compositional convection, concerned with the inner core solidification, leads to  increase of convection intensity in the core. 
    In units of its critical value  the Rayleigh number   increases in factor 20 that corresponds to increase of the intensity of the geomagnetic field times  $(20)^{ 1\over 3}\approx 2.7$ \cite{ChAu2006}. Moreover, influence  of the inner core on the magnetic field generation can be even smaller. The reason is that compositional convection generates magnetic field located deeper in the liquid core, and as a result, intensity of the magnetic field  at the surface of the planet is weaker than in the case of the thermal convection.
     In some sense our result confirms recent results \cite{Driscoll2016} that switch on of the compositional convection  does not change generation of the magnetic field essentially, as it was supposed earlier \cite{Olson2013}.  
 However the absolute values of the Rayleigh numbers are still under consideration, and their  comparison for the compositional and thermal convection models is a tricky procedure, the time evolution of these parameters is not so questionable: the considered above scenarios predict increase of the geomagnetic reversals frequency with the inner core growth in agreement  with observations.

 It is worth noting that our  analysis includes two independent physical data: concerned with the thermal evolution of the core, and evolution of the length of the day, based on  observations. Of course, these data are  interrelated, and the more complicated physical model is required for description. The hint is that identification of the break in the curve of length of the day  with the origin of the inner core simplifies the general evolution scenario for the considered model. In this case  the young inner core is more likely.


 \bibliographystyle{mhd}

\newcommand{\noopsort}[1]{}

\vfill\eject

\end{document}